\documentclass[aps,prd,reprint,superscriptaddress,floatfix,nofootinbib]{revtex4-1}

\usepackage[T1]{fontenc}         
\usepackage[utf8]{inputenc}      
\usepackage{graphicx}            
\usepackage{xcolor}              
\usepackage{mathtools}           
\usepackage{amssymb}             
\usepackage{lmodern}             
\usepackage{bm}                  
\usepackage{siunitx}             
\usepackage[final]{microtype}    
\usepackage{hyperref}            

\graphicspath{{./Graphics/}}

\hypersetup{%
	pdftitle = {Baryon number fluctuations in the QCD phase diagram from Dyson-Schwinger equations},
	pdfauthor = {Philipp Isserstedt, Michael Buballa, Christian S. Fischer, Pascal J. Gunkel},
	bookmarksopen = true,
	colorlinks = true,
	linkcolor = red!50!black,
	citecolor = blue!50!black,
	urlcolor = blue!50!black
}

\newcommand*{\+}{\hspace*{.08335em}}
\newcommand*{\beq}{\begin{equation}}
\newcommand*{\eeq}{\end{equation}}
\newcommand*{\Nf}{N_{\textup{f}}}
\newcommand*{\Nc}{N_{\textup{c}}}
\newcommand*{\Tc}{T_{\textup{c}}}
\newcommand*{\OO}{\textup{O}}
\newcommand*{\SU}{\textup{SU}}
\newcommand*{\ZZ}{\textup{Z}}
\newcommand*{\dd}{\textup{d}}
\newcommand*{\ii}{\textup{i}}
\newcommand*{\upB}{\textup{B}}
\newcommand*{\upS}{\textup{S}}
\newcommand*{\upQ}{\textup{Q}}
\newcommand*{\upu}{\textup{u}}
\newcommand*{\upd}{\textup{d}}
\newcommand*{\ups}{\textup{s}}
\newcommand*{\ZT}{Z_{\textup{T}}}
\newcommand*{\ZL}{Z_{\textup{L}}}
\newcommand*{\muu}{\mu_{\upu}}
\newcommand*{\mud}{\mu_{\upd}}
\newcommand*{\mus}{\mu_{\ups}}
\newcommand*{\muB}{\mu_{\upB}}
\newcommand*{\muS}{\mu_{\upS}}
\newcommand*{\muQ}{\mu_{\upQ}}
\newcommand*{\bbN}{\mathbb{N}}
\newcommand*{\bbZ}{\mathbb{Z}}

\renewcommand*{\vec}[1]{\bm{#1}}

\DeclareMathOperator{\Tr}{Tr}
\DeclareMathOperator*{\argmax}{argmax}
\DeclarePairedDelimiterX{\expval}[1]{\langle}{\rangle}{#1}

\DeclareMathOperator{\sumint}{\mathchoice%
  {\ooalign{\raisebox{.08\height}{\scalebox{.85}{$\displaystyle\sum$}}\cr\hidewidth$\displaystyle\int$\hidewidth\cr}}%
  {\ooalign{\raisebox{.24\height}{\scalebox{.6}{$\textstyle\sum$}}\cr\hidewidth$\textstyle\int$\hidewidth\cr}}%
  {\ooalign{\raisebox{.24\height}{\scalebox{.6}{$\scriptstyle\sum$}}\cr$\scriptstyle\int$\cr}}%
  {\ooalign{\raisebox{.24\height}{\scalebox{.6}{$\scriptstyle\sum$}}\cr$\scriptstyle\int$\cr}}%
}

\DeclareSIUnit{\eV}{\electronvolt}

\AtBeginDocument{\sisetup{separate-uncertainty}}

\begin{document}

\title{Baryon number fluctuations in the QCD phase diagram\texorpdfstring{\\}{ }from Dyson--Schwinger equations}

\author{Philipp Isserstedt}
\email{philipp.isserstedt@physik.uni-giessen.de}
\affiliation{%
	Institut f\"{u}r Theoretische Physik, %
	Justus-Liebig-Universit\"{a}t Gie\ss{}en, %
	35392 Gie\ss{}en, %
	Germany%
}

\author{Michael Buballa}
\email{michael.buballa@physik.tu-darmstadt.de}
\affiliation{%
	Theoriezentrum, %
	Institut f\"{u}r Kernphysik, %
	Technische Universit\"{a}t Darmstadt, %
	64289 Darmstadt, %
	Germany%
}

\author{Christian S.~Fischer}
\email{christian.fischer@theo.physik.uni-giessen.de}
\affiliation{%
	Institut f\"{u}r Theoretische Physik, %
	Justus-Liebig-Universit\"{a}t Gie\ss{}en, %
	35392 Gie\ss{}en, %
	Germany%
}

\author{Pascal J.~Gunkel}
\email{pascal.gunkel@physik.uni-giessen.de}
\affiliation{%
	Institut f\"{u}r Theoretische Physik, %
	Justus-Liebig-Universit\"{a}t Gie\ss{}en, %
	35392 Gie\ss{}en, %
	Germany%
}

\begin{abstract}
We present results for fluctuations of the baryon number for QCD at nonzero temperature and chemical potential. These are extracted from solutions to a coupled set of truncated Dyson--Schwinger equations for the quark and gluon propagators of Landau gauge QCD with $\Nf = 2 + 1$ quark flavors that has been studied previously. We discuss the changes of fluctuations and ratios thereof up to fourth order for several temperatures and baryon chemical potential up to and beyond the critical endpoint. In the context of preliminary STAR data for the skewness and kurtosis ratios, the results are compatible with the scenario of a critical endpoint at large chemical potential and slightly offset from the freeze-out line. We also discuss the caveats involved in this comparison.
\end{abstract}

\maketitle 

\section{\label{sec:intro}Introduction}

Extracting the location of a putative critical endpoint (CEP) of QCD from heavy-ion 
collisions is one of the major goals of the Beam Energy Scan (BES) program 
\cite{Aggarwal:2010cw,STAR:whitepaper} at the Relativistic Heavy Ion Collider (RHIC) 
at Brookhaven National Laboratory and the future Compressed Baryonic Matter (CBM) 
experiment \cite{Friman:2011zz} at the Facility for Antiproton and Ion Research (FAIR).

Theoretically it is by no means clear that such a critical endpoint exists. At zero chemical
potential, there is firm evidence from lattice QCD for an analytic crossover from a low-temperature 
phase characterized by chiral symmetry breaking to a high-temperature (partially) chirally restored phase
\cite{Aoki:2006we,Aoki:2009sc,Borsanyi:2010bp,Bazavov:2011nk,Bhattacharya:2014ara,Bazavov:2014pvz}. 
The corresponding pseudocritical temperature has been localized at $\Tc \approx \SI{156}{\mega\eV}$ within 
a definition-dependent range of several MeV
\cite{Borsanyi:2010bp,Bazavov:2011nk,Bellwied:2015rza,Bazavov:2018mes}.
However, the situation is much less clear at 
(real) chemical potential, where lattice calculations are hampered by the notorious fermion sign problem.
Model calculations suggest that the continuous crossover becomes steeper with increasing chemical potential 
and finally merges into a second-order CEP followed by a region of a first-order 
phase transition at large chemical potential \cite{Asakawa:1989bq,Stephanov:1998dy,Stephanov:1999zu,
Fukushima:2003fw,Megias:2004hj,Ratti:2005jh,Schaefer:2007pw,Skokov:2010sf,Skokov:2010wb,Herbst:2010rf}; 
see, e.g., Refs.~\cite{Drews:2016wpi,Fukushima:2017csk} for review articles. This notion is supported by results 
from Dyson-Schwinger equations
\cite{Qin:2010nq,Fischer:2012vc,Fischer:2014ata,Eichmann:2015kfa,Gao:2016qkh}, see Ref.~\cite{Fischer:2018sdj} 
for a recent review. 

In order to put these theoretical ideas to the test in experiments, observables have been
identified that are
capable to deliver signals of the CEP. Provided the freeze-out in heavy-ion collisions is sufficiently
close to the phase boundary, fluctuations of conserved charges (baryon number, strangeness, and 
electric charge) are expected to provide this information
\cite{Stephanov:1998dy,Stephanov:1999zu,Jeon:1999gr,Jeon:2000wg,Asakawa:2000wh,Koch:2005vg,Ejiri:2005wq}.
Various ratios of cumulants of these conserved quantities can be extracted from experiment in 
event-by-event analyses and compared to corresponding ratios of fluctuations that can be
determined in theoretical calculations, see, e.g., Refs.~\cite{Luo:2017faz,Bzdak:2019pkr} for reviews.
Lattice-QCD results for fluctuations and correlations at zero \cite{Cheng:2008zh,Borsanyi:2011sw,Bazavov:2012jq,Borsanyi:2013hza,Bellwied:2015lba}
and small chemical potential \cite{Bazavov:2017tot,Borsanyi:2018grb}
are available, but need to be extended toward higher chemical potential. 

In hadron resonance gas (HRG) approaches and refined effective models, such as the Polyakov-loop enhanced 
Nambu-Jona-Lasinio model and the Polyakov-loop quark-meson model (PQM), a wealth of interesting
results on fluctuations have been obtained already, see, e.g.,
Refs.~\cite{Skokov:2010wb,Skokov:2010uh,Karsch:2010hm,Schaefer:2011ex,Morita:2014fda,Morita:2014nra,
Fu:2015naa,Fu:2015amv,Fu:2016tey,Shao:2017yzv,Fu:2018swz,Szymanski:2019yho,Yin:2019ebz} and
references therein. Concerning the Yang-Mills sector, these models rely on the Polyakov loop potential
that couples aspects of confinement to the chiral dynamics, however without backcoupling. Thus, gluons
are no active degrees of freedom (d.o.f.) and their reaction to the medium can neither be studied nor directly taken into account.

A different approach is possible in functional methods. In a series of works
\cite{Fischer:2012vc,Fischer:2014ata,Fischer:2014vxa,Eichmann:2015kfa} a coupled system of 
Dyson--Schwinger equations (DSEs) for the quark and gluon propagators has been considered 
and the physics of the Columbia plot \cite{Brown:1990ev}
has been explored. Results for QCD with heavy quarks and at physical quark masses ($\Nf=2+1$ and $\Nf=2+1+1$)
but zero chemical potential agree with corresponding lattice results, see Ref.~\cite{Fischer:2018sdj}
for an overview. A critical endpoint has been found at
$\bigl( T^\textup{CEP}, \, \muB^\textup{CEP} \bigr) = (117,488) \, \si{\mega\eV}$
that corresponds to a ratio $\muB^\textup{CEP} / \, T^\textup{CEP} \approx 4.2$, i.e., large chemical
potential. In this work we will use this framework to explore cumulants and ratios thereof along the
crossover line from $\muB=0$ to and beyond the CEP.
We thereby improve previous results for fluctuations calculated in the DSE framework of Ref.~\cite{Xin:2014ela}, where backcoupling effects have not been
taken into account. In particular, we discuss ratios involving the skewness and kurtosis and compare
our results with  preliminary data from the STAR collaboration extracted from the BES at RHIC.   

This work is organized as follows: In Sec.~\ref{sec:fluctuations}, we detail  
our method to extract fluctuations from the quark propagator and derivatives 
thereof. In Sec.~\ref{sec:dse}, we summarize the truncation scheme of the DSEs and discuss
the (slight) changes as compared to previous works \cite{Fischer:2012vc,Fischer:2014vxa}. In 
Sec.~\ref{sec:results}, we present our results and finally conclude in Sec.~\ref{sec:summary}.
 
\section{\label{sec:fluctuations}fluctuations}

In $\Nf=2+1$ flavor QCD, there is a conserved charge for each quark flavor controlled by the
three quark chemical potentials $\muu$, $\mud$, and $\mus$. The quantities under
study in the present work are fluctuations of these conserved charges, i.e., higher-order derivatives of
the grand-canonical potential
\beq
	\label{eq:omega_general}
	\Omega = -\frac{T}{V} \log \mathcal{Z}(T,\muu,\mud,\mus)
\eeq
with respect to the quark chemical potentials. Here, $\mathcal{Z}$ is the partition function
of QCD, $T$ the temperature, and $V$ the volume of the system. The fluctuations are then
included in%
\footnote{%
	We usually suppress the arguments of the fluctuations. However, one has to keep
	in mind that they are functions of temperature and all chemical potentials,
	i.e., $\chi_{ijk}^{\upu\upd\ups} \equiv \chi_{ijk}^{\upu\upd\ups}(T,\muu,\mud,\mus)$.
	If a subscript is vanishing, it is omitted together with its superscript counterpart,
	e.g., $\chi_2^\upu \equiv \chi_{200}^{\upu\upd\ups}$. 
}
\beq
	\label{eq:fluctuations}
	\chi_{ijk}^{\upu\upd\ups}
	=
	-\frac{1}{T^{4-(i+j+k)}}
	\frac{\partial^{\+i+j+k} \+ \Omega}{\partial \muu^i \+ \partial \mud^j \+ \partial \mus^k}
\eeq
with $i,\+ j,\+ k \in \bbN_0$. The quark chemical potentials are related to the ones for
baryon number (B), strangeness (S), and electric charge (Q) via
\begin{align}
	\label{eq:mu_relations_u}
	\muu &= \frac{1}{3} \+ \muB + \frac{2}{3} \+ \muQ \, ,
	\\[.25em]
	\label{eq:mu_relations_d}
	\mud &= \frac{1}{3} \+ \muB - \frac{1}{3} \+ \muQ \, ,
	\\[.25em]
	\label{eq:mu_relations_s}
	\mus &= \frac{1}{3} \+ \muB - \frac{1}{3} \+ \muQ - \muS \, .
\end{align}
With these relations one finds for example the second-order baryon number fluctuation
\beq
	\label{eq:chi_2_B}
	\begin{aligned}
		\chi_2^\upB &= -\frac{1}{T^2} \frac{\partial^2 \+ \Omega}{\partial \muB^2}
		\\[.25em]
		&=
		\frac{1}{9}
		\left[
			\chi_2^\upu + \chi_2^\upd + \chi_2^\ups
			+ 2 \left( \chi_{11}^{\upu\ups} + \chi_{11}^{\upd\ups} + \chi_{11}^{\upu\upd} \right)
		\right]
	\end{aligned}
\eeq
in terms of quark degrees of freedom. Other fluctuations can be determined analogously.

Ratios of fluctuations in baryon number, electric charge, and strangeness are particularly interesting 
since they are equal to corresponding ratios of cumulants that can be extracted from experimental 
quantities accessible in event-by-event analyses of heavy-ion collisions, see the review articles
\cite{Asakawa:2015ybt,Luo:2017faz,Bzdak:2019pkr} for more details. Interesting ratios related
to the baryon number are
\beq
	\label{eq:chi_ratios}
	\begin{gathered}
		\frac{\chi_4^\upB}{\chi_2^\upB} = \kappa_\upB \+ \sigma_\upB^2 \, ,
		\qquad
		\frac{\chi_3^\upB}{\chi_2^\upB} = S_\upB \+ \sigma_\upB \, ,
		\\[.5em]
		\frac{\chi_1^\upB}{\chi_2^\upB} = \frac{M_\upB}{\sigma_\upB^2} \, ,
	\end{gathered}
\eeq
where $\kappa_\upB$, $\sigma_\upB^2$, $S_\upB$, and $M_\upB$ denote the kurtosis, variance,
skewness, and mean of the net-baryon distribution, respectively. Ratios of fluctuations are 
a suitable tool to explore the phase diagram of QCD since they are sensitive to phase transitions \cite{Stephanov:1998dy,Stephanov:1999zu,Asakawa:2000wh,Jeon:2000wg,Stephanov:2004wx,Koch:2005vg,
Ejiri:2005wq,Friman:2011pf}. At the critical endpoint, the correlation length $\xi$ diverges 
(at least for infinite volume) and $\chi_2^\upB \sim \xi^c$ with $c > 0$. 

From the first BES at RHIC, the STAR collaboration extracted results for the net-proton
number fluctuations $M_\textup{P}$, $\sigma^2_\textup{P}$, $S_\textup{P}$, and $\kappa_\textup{P}$
\cite{Aggarwal:2010wy,Luo:2015ewa}, which can be used as a proxy for fluctuations of the net-baryon
number. The data suggest a number of interesting tendencies that are drastically different from
results of HRG model calculations, but agree with results from lattice QCD
\cite{DElia:2016jqh,Bazavov:2017tot,Borsanyi:2018grb} obtained for small baryon chemical potential.
As already mentioned in the introduction, it is the purpose of this paper to provide theoretical results
for larger chemical potential in the framework of functional approaches to QCD.  

In the present work, we consider the lowest-order fluctuations with $1 \leq i + j + k \leq 4$ and 
determine them via the quark number densities.
We start with the grand-canonical potential expressed as a functional of the propagators of QCD.
Consequently, $\Omega$ contains contributions from quarks, gluons, and ghosts. Since the latter are
only weakly chemical-potential dependent, we neglect their contributions and arrive at
\cite{Cornwall:1974vz}
\beq
	\label{eq:Omega}
	\Omega
	=
	-\frac{T}{V}
	\left(
		\Tr\log \frac{S^{-1}}{T}
		- \Tr\bigl[ \openone - S_0^{-1} S \bigr]
		+ \Phi_\textup{int}[S]
	\right)
\eeq
with $S$ being the dressed quark propagator with flavor, color, Dirac, and momentum degrees
of freedom; $S_0$ denotes its bare counterpart. The trace has to be taken in the functional sense over
flavor, color, Dirac, and momentum space, and the interaction functional $\Phi_\textup{int}$ contains
all two-particle irreducible diagrams with respect to $S$. Note that Eq.~\eqref{eq:Omega} is nothing
but the two-particle irreducible effective action evaluated at the stationary point and therefore
$\delta \+ \Omega \+ / \+ \delta S = 0$.

The quark number densities then read\footnote{%
	We work in four-dimensional Euclidean space-time with Hermitian gamma matrices obeying
	$\{ \gamma_\mu, \gamma_\nu \} = 2 \+ \delta_{\mu\nu}$; $\mu,\nu \in \{ 1, 2, 3, 4 \}$.
	Our choice for the heat bath vector is $u=(u_4,\vec{u})=(1, \vec{0})$.
}
\beq
	\label{eq:rho_f}
	n_f
	= -\frac{\partial \+ \Omega}{\partial \mu_f}
	= -\Nc \+ Z_2^f \+ \sumint_q \Tr \bigl[ \gamma_4 \+ S_f(q) \bigr] \, ,
\eeq
where $f \in \{ \upu, \upd, \ups \}$ labels the flavor, $\Nc=3$ denotes the number of colors,
$Z_2^f$ is the quark wave function renormalization constant,
and $q = (\omega_q, \vec{q})$ with fermionic
Matsubara frequencies $\omega_q = (2 \+ \ell_q + 1) \+ \pi T$; $\ell_q \in \bbZ$.
The Matsubara sum as well as the three-momentum integration is 
abbreviated by $\sumint_q \equiv T \sum_{\ell_q \in \+ \bbZ} \int \dd^3 \vec{q} \+ / \+ (2\pi)^3$,
and the remaining trace has to be evaluated in Dirac space.
The quantity $S_f(q)$ denotes the dressed quark propagator at nonzero temperature
and chemical potential with only Dirac and momentum degrees of freedom left. Equation \eqref{eq:rho_f} can also be written as an expectation value,
$n_f = \expval{\psi^\dagger \psi}_f = \expval{\bar{\psi} \+ \gamma_4 \+ \psi}_f$,
and its gauge invariance as a density of a global charge is guaranteed by the Landau--Khalatnikov--Fradkin
transformations \cite{Landau:1955zz,*Fradkin:1955jr,Zumino:1959wt}.

The quark number densities need to be evaluated using a very large number
of Matsubara frequencies in order to obtain stable results. In addition, with nonperturbative
propagators and a numerical cutoff in the three-momentum integral, this expression needs to be
regularized. To this end, we employ the subtraction scheme used in Refs.~\cite{Gao:2015kea,Gao:2016qkh}
that is an Euclidean version of the contour-integration technique for Matsubara sums
\cite{Kapusta:2006pm}.  The regularized quark number density is given by 
\beq
	\label{eq:rho_f_reg}
	n_f^\textup{reg} = -\Nc \+ Z_2^f \int \frac{\dd^3 \vec{q}}{(2 \pi)^3} \+ K_f(\vec{q})
\eeq
with
\beq
	\label{eq:rho_f_reg_kernel}
	\begin{aligned}
		K_f(\vec{q})
		&=
		T \sum_{\ell_q \in \+ \bbZ} \Tr\bigl[ \gamma_4 \+ S_f(\omega_q, \vec{q}) \bigr]
		\\[.25em]
		& \phantom{=\;}
		- \frac{1}{2\pi} \int_{-\infty}^\infty \dd q_4
		\Tr\bigl[ \gamma_4 \+ S_f(q_4, \vec{q}) \bigr] \, .
	\end{aligned}
\eeq
The last term does not depend explicitly on temperature or chemical potential and is known as
a ``vacuum contribution'' in the literature \cite{Kapusta:2006pm}. We verified that its subtraction
leads to cutoff-independent results.

Having the quark number densities at hand, the fluctuations are obtained by higher-order
derivatives of these densities. For example, the second-order up quark fluctuation reads
\beq
	\label{eq:chi_2_u}
	\chi_2^\upu = \frac{1}{T^2} \frac{\partial \+ n_\upu^\textup{reg}}{\partial \muu}
\eeq
and other quantities are obtained analogously. 

\section{\label{sec:dse}Dyson--Schwinger equations}

In the following, we briefly summarize the functional framework used to determine the temperature
and chemical potential dependent dressed quark propagator needed to determine the quark densities
of Eq.~\eqref{eq:rho_f}. To this end, we solve a set of truncated Dyson--Schwinger equations. In contrast
to previous works on fluctuations in the DSE framework \cite{Xin:2014ela,Xu:2015jwa} we take the 
back reaction of the quarks onto the Yang--Mills sector explicitly into account. This establishes
a temperature and chemical-potential dependence of the gluon controlled by QCD dynamics rather than
simple modelling. Furthermore, this allows for explicit control over the quark-flavor dependence
of all results. Our framework evolved from the quenched case 
\cite{Fischer:2009wc,Fischer:2010fx}, to $\Nf=2$ \cite{Fischer:2011pk,Fischer:2011mz,Fischer:2012vc}, and 
finally to $\Nf=2+1$ and $\Nf=2+1+1$ quark flavors with physical quark masses \cite{Fischer:2012vc,Fischer:2014ata}. 

With $\OO(4)$ symmetry broken to $\OO(3)$ due to the heat bath,
the dressed inverse quark propagator $S_f^{-1}$ for a flavor $f$ at nonzero temperature $T$
and quark chemical potential $\mu_f$ is given by
\beq
	\label{eq:Sinv_dressed}
	S_f^{-1}(p)
	=
	\ii \+ (\omega_p + \ii \+ \mu_f) \+ \gamma_4 \+ C_f(p)
	+ \ii \+ \vec{\gamma} \cdot \vec{p} \+ A_f(p)
	+ B_f(p)
\eeq
with momentum $p = (\omega_p, \vec{p})$ and fermionic Matsubara frequencies
$\omega_p = (2 \+ \ell_p + 1) \+ \pi T$, $\ell_p \in \bbZ$. The dressing functions $C_f$,
$A_f$, and $B_f$ depend on momentum and contain all nonperturbative information. Furthermore,
they depend on temperature and chemical potential. The corresponding bare quark propagator reads
\beq
	\label{eq:Sinv_bare}
	S_{0,f}^{-1}(p)
	=
	Z_2^f
	\+
	\bigl(
		\+ \ii \+ (\omega_p + \ii \+ \mu_f) \+ \gamma_4
		+ \ii \+ \vec{\gamma} \cdot \vec{p} \+
		+ Z_m^f \+ m_f \+
	\bigr)
\eeq
with $Z_m^f$ denoting the quark mass renormalization constant,
and $m_f$ is the renormalized current quark mass. In principle, there is a fourth
Dirac structure $\gamma_4 \+ \vec{\gamma} \cdot \vec{p}$ contributing to the inverse quark
propagator. However, its contribution is negligible \cite{Contant:2017gtz} and therefore not
considered in this work.

Since we work in Landau gauge, the gluon propagator is purely transverse with respect to its
four-momentum $k = (\omega_k, \vec{k})$, where $\omega_k = 2 \+ \ell_k \pi T$ ($\ell_k \in \bbZ$)
are bosonic Matsubara frequencies. However, due to the presence of the heat bath, the transverse
space splits into two parts, and the dressed gluon propagator reads
\beq
	\label{eq:D_dressed}
	D_{\mu\nu}(k)
	=
	P_{\mu\nu}^\textup{T}(k) \+ \frac{\ZT(k)}{k^2}
	+ P_{\mu\nu}^\textup{L}(k) \+ \frac{\ZL(k)}{k^2} \,,
\eeq
with projectors 
\begin{align}
	\label{eq:hb_projectors}
	P_{\mu\nu}^\textup{T}(k)
	&=
	(1 - \delta_{4\mu})
	\,
	(1 - \delta_{4\nu})
	\left( \delta_{\mu\nu} - \frac{k_\mu k_\nu}{\vec{k}^2} \right) ,
	\\[.25em]
	P_{\mu\nu}^\textup{L}(k)
	&=
	\delta_{\mu\nu} - \frac{k_\mu k_\nu}{k^2} - P_{\mu\nu}^\textup{T}(k) \, .
\end{align}

\begin{figure}[t]
	\centering%
	\includegraphics[scale=1.1]{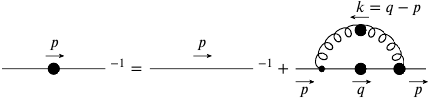}%
	\vspace*{-0.5em}%
	\caption{\label{fig:qDSE}%
		The DSE for the quark propagator. Large filled circles denote dressed quantities;
		solid and wiggly lines represent quarks and gluons, respectively. There is a separate
		DSE  for each quark flavor.
	}
	\vspace*{1em}%
	\includegraphics[scale=1.05]{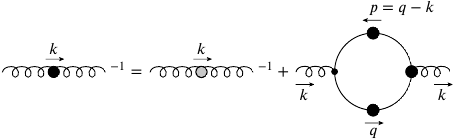}%
	\vspace*{-0.5em}%
	\caption{\label{fig:gDSE}%
		The DSE for the gluon propagator. The gray circle denotes the bare gluon propagator together
		with all diagrams with no explicit quark content. The flavor sum for the quark loop diagram
		is implicit;  we consider $\Nf=2+1$ quark flavors.
	}
\end{figure}

The dressed quark and the dressed gluon propagator each satisfy a Dyson--Schwinger equation
that read
\begin{gather}
	\label{eq:DSE_full}
	S_f^{-1}(p) = S_{0,f}^{-1}(p) + \Sigma_f(p) \, ,
	\\[.25em]
	D_{\mu\nu}^{-1}(k) = \bigl[ D_{\mu\nu}^\textup{YM}(k) \bigr]^{-1} + \Pi_{\mu\nu}(k) \,.
\end{gather}
The symbol $D_{\mu\nu}^\textup{YM}$ denotes the sum of the inverse bare gluon propagator and all
diagrams with no explicit quark content. The quark self-energy $\Sigma_f$ and the gluon self-energy
from the quark loop $\Pi_{\mu\nu}$ are given by 
\begin{align}
	\label{eq:quark_selfenergy}
	\Sigma_f(p)
	&=
	C_\textup{F} \+ g^2 \+ \frac{Z_2^f}{\widetilde{Z}_3} \+ \sumint_q
	\gamma_\mu \+ D_{\mu\nu}(k) \+ S_f(q) \+ \Gamma_\nu^f(q,p;k) \, ,
	\\
	\label{eq:quark_loop}
	\Pi_{\mu\nu}(k)
	&=
	-\frac{g^2}{2} \+ \sum_f \+ \frac{Z_2^f}{\widetilde{Z}_3} \+
	\sumint_q \Tr \bigl[ \gamma_\mu \+ S_f(q) \+ \Gamma_\nu^f(q,p;k) \+ S_f(p) \bigr]
\end{align}
with $f \in \{ \upu, \upd, \ups \}$. The equations are shown diagrammatically in Figs.~\ref{fig:qDSE}
and \ref{fig:gDSE}. Note that different flavors are nontrivially coupled
through the quark loop $\Pi_{\mu\nu}$. Furthermore, $k = q - p$ and $p = q - k$ in the quark
and gluon DSE, respectively, $\widetilde{Z}_3$ is the ghost renormalization constant, and
$\Gamma_\nu^f$ denotes the dressed quark-gluon vertex. For the coupling we use
$\alpha = g^2 / \+ (4\pi) = 0.3$ (see Refs.~\cite{Fischer:2009wc,Fischer:2010fx} for details) 
and $C_\textup{F} = (\Nc^2 - 1) \+ / \+ (2 \Nc)$ is the $\SU(\Nc)$ quadratic Casimir factor
in the fundamental representation that stems from the color trace.

In order to determine the gluon propagator at nonzero temperature and chemical potential,
an efficient approximation that has been used in the literature is to replace the Yang--Mills
part of the equation, $D_{\mu\nu}^\textup{YM}$, by quenched temperature-dependent lattice data
\cite{Fischer:2010fx,Maas:2011ez}. This approximation misses implicit quark-loop effects in the
Yang--Mills self-energies, which are subleading in a $1 \+ / \+ \Nc$ expansion as
compared to the explicit quark loop $\Pi_{\mu\nu}$. At zero temperature, the effects of this
approximation can be estimated using the framework of Ref.~\cite{Fischer:2003rp} and are found to
be well below the five-percent level. This strategy has been used in
Refs.~\cite{Fischer:2014ata,Fischer:2014vxa,Eichmann:2015kfa} to determine the location of the
critical endpoint and will be adopted also in this work. 

Eventually, the quark-gluon vertex is the last quantity that needs to be specified to obtain
a closed system of equations. We use the following ansatz (see Ref.~\cite{Fischer:2014ata} for more
details): The leading term of the Ball--Chiu vertex construction \cite{Ball:1980ay} is multiplied by
a phenomenological vertex dressing function $\Gamma$ that accounts for non-Abelian effects and the
correct logarithmic running of the propagators in the ultraviolet. The resulting equations are
\beq
	\label{eq:vertex_ansatz}
	\begin{aligned}
		\frac{1}{\widetilde{Z}_3} \+ \Gamma_\nu^f(q,p;k)
		&=
		\Gamma(x) \+ \biggl[ \+ \delta_{4\nu} \+ \frac{C_f(q) + C_f(p)}{2}
		\\[.25em]
		& \phantom{=\;}
		+ (1 - \delta_{4\nu}) \+ \frac{A_f(q) + A_f(p)}{2} \+ \biggr] \+ \gamma_\nu
	\end{aligned}
\eeq
and
\beq
	\label{eq:vertex_dressing}
	\Gamma(x)
	=
	\frac{d_1}{d_2 + x}
	+
	\frac{1}{1 + \Lambda^2 / \+ x}
	\left(
		\frac{\alpha \beta_0}{4\pi}
		\log \! \left( 1 + \frac{x}{\Lambda^2} \right)
	\right)^{2\delta},
\eeq 
where $\delta = -9 \Nc \+ / \+ (44 \Nc - 8 \Nf)$ is the anomalous dimension of the vertex and
$\beta_0 = (11 \Nc - 2 \Nf) \+ / \+ 3$. The squared momentum argument of $\Gamma$ is $x = k^2$ 
in the quark self-energy but
$x = p^2 + q^2$ in the quark loop. This is necessary to maintain multiplicative renormalizability
of the gluon DSE \cite{Fischer:2003rp}. Note that the vertex ansatz includes effects from nonzero
temperature and chemical potential, as the full vertex certainly would, since the terms from the
Ball--Chiu construction involve the quark dressing functions. The parameters $d_2 = \SI{0.5}{\giga\eV\squared}$
and $\Lambda = \SI{1.4}{\giga\eV}$ are fixed to match the scales in the quenched gluon
propagator from the lattice.

\begin{figure*}[t]
	\centering%
	\includegraphics[trim=0 2mm 0 0]{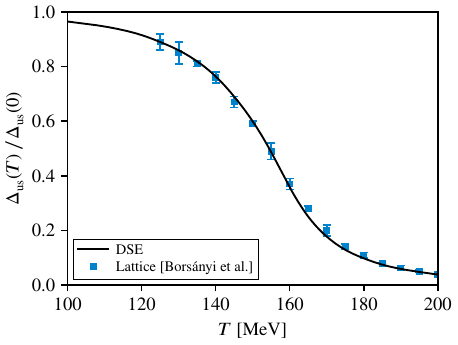}%
	\hspace*{2em}%
	\includegraphics[trim=0 2.05mm 0 0]{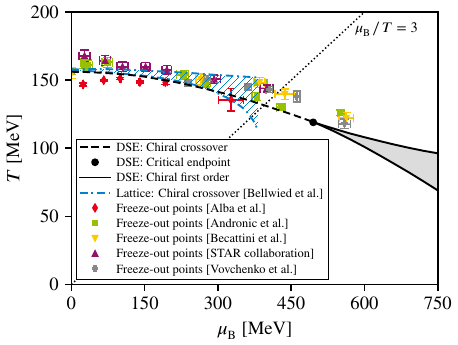}%
	\caption{\label{fig:condensate_phase_diagram}
		Left:
		Subtracted quark condensate normalized to its vacuum value as a function of
		temperature at vanishing chemical potential compared to the continuum-extrapolated
		lattice result of Ref.~\cite{Borsanyi:2010bp}.
		Right:
		Our result for the phase diagram for $\Nf = 2 + 1$ quark flavors compared to 
		freeze-out points from heavy-ion collisions extracted by different methods/groups
		\cite{Alba:2014eba,Becattini:2016xct,Vovchenko:2015idt,Adamczyk:2017iwn,Andronic:2016nof,Andronic:2017pug}.
		Shown is also the region of the chiral crossover from lattice QCD (blue band) \cite{Bellwied:2015rza} 
		(see also Ref.~\cite{Bazavov:2018mes}).
	}
\end{figure*}

\subsection*{Quark masses, vertex strength, and chemical potentials}

The remaining value of the vertex strength parameter $d_1$ as well as the quark masses
$m_{\upu,\upd,\ups}$ are fixed using lattice results for the subtracted quark condensate.
The quark condensate is given by
\beq
	\expval{\bar{\psi}\psi}_f = -\Nc \+ Z_2^f Z_m^f \+ \sumint_q \Tr \bigl[ S_f(q) \bigr]
\eeq
for each quark flavor $f$. It is plagued by a quadratic divergence for all flavors with nonzero 
quark masses and needs to be regularized. This can be accomplished by the difference
\beq
	\Delta_{\upu\ups}
	=
	\expval{\bar{\psi} \psi}_\upu
	-
	\frac{m_\upu}{m_\ups} \+ \expval{\bar{\psi} \psi}_\ups \, ,
\eeq
which defines the subtracted quark condensate. It is an order parameter for chiral symmetry 
breaking and may be used to define the pseudocritical temperature, see Eq.~(\ref{eq:Tc_definition})
below. We adapt the vertex strength parameter $d_1 = \SI{8.49}{\giga\eV\squared}$ such that 
the pseudocritical temperature found on the lattice 
\cite{Borsanyi:2010bp,Bazavov:2011nk,Bellwied:2015rza,Bazavov:2018mes} is reproduced. We work in the
isospin-symmetric limit of equal up and down quark masses, $m_\upu = m_\upd$. In the high-temperature 
phase, $\Delta_{\upu\ups}$ is mainly controlled by these masses and
we adapt their values to match the lattice results. Finally we fix the up-to-strange quark mass ratio of
$m_\ups \+ / \+ m_\upu = 25.7$ using results for the pion and kaon masses in vacuum obtained 
from the Bethe--Salpeter formalism developed in Ref.~\cite{Heupel:2014ina}. This results in 
$m_\upu(\zeta) = \SI{0.8}{\mega\eV}$ and $m_\ups(\zeta) = \SI{20.56}{\mega\eV}$ at a renormalization point
of $\zeta=\SI{80}{\giga\eV}$, i.e., far in the perturbative regime.
Note that the values of $d_1$ and $m_{\upu,\ups}$ are slightly different from the ones 
reported in Ref.~\cite{Fischer:2014ata} because we employ a slightly lighter strange quark mass and an
Pauli--Villars regulator%
\footnote{%
	This amounts to $D_{\mu\nu}(k) \to D_{\mu\nu}(k) \+ / \+ (1 + k^2 / \+ \Lambda_\textup{PV}^2)$
	in Eq.~\eqref{eq:quark_selfenergy}.
	For the Pauli--Villars scale we use $\Lambda_\textup{PV} = \SI{200}{\giga\eV}$.
}
for the quark DSE rather than a hard cutoff. 

In principle, the chemical potentials should be adjusted in order to implement strangeness neutrality
as encountered in a heavy-ion collision. This is done by an appropriate dependence of $\muQ$ and $\muS$
on $\muB$. For temperatures around $\SI{150}{\mega\eV}$,
the leading-order result from lattice QCD is $\muQ \approx -0.02 \, \muB$ while $\muS \approx 0.2 \, \muB$
\cite{Borsanyi:2013hza,Bazavov:2012vg}. Thus, to a good approximation we choose $\muu = \mud$. 
Furthermore, it has been checked within the framework of DSEs that values for the strange quark chemical 
potential between $\mus=0$ and $\mus=\mud$
hardly affects the location of the CEP \cite{Welzbacher:2016}. Thus, for the purpose of this work we 
choose $\mus=0$ and, the baryon chemical potential is then given by $\muB = 3 \+ \muu$.%
\footnote{In Ref.~\cite{Fu:2018qsk}, the impact of strangeness neutrality on thermodynamic observables is
studied. Since at large chemical potentials quantitative corrections of the order of 20\% for some
thermodynamic quantities have been found, which may also affect fluctuations, we strive to implement
strangeness neutrality in future work.}

\section{\label{sec:results}Results}

\subsection{\label{sec:results:phase_diagram}Phase diagram}

\begin{figure*}[t]
	\centering%
	\includegraphics[trim=0 2mm 0 0]{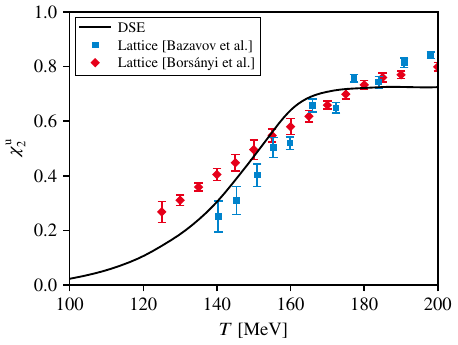}%
	\hspace*{2em}%
	\includegraphics[trim=0 2mm 0 0]{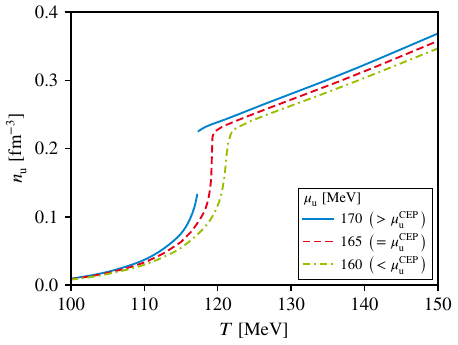}%
	\caption{\label{fig:chi_2_u}%
		Left:
		Second-order up/down quark fluctuation at vanishing chemical potential. The lattice data 
		are taken from Refs.~\cite{Bazavov:2011nk} and \cite{Borsanyi:2011sw}, respectively.
		Right:
		Up-quark number density in the vicinity of the CEP for three different chemical potentials.
	}%
\end{figure*}

Before we discuss fluctuations, we present our updated result for the QCD phase diagram 
with $\Nf = 2 + 1$ quark flavors, which closely resembles the one already published in 
Refs.~\cite{Fischer:2014ata,Fischer:2018sdj}. Overall, changes due to the slightly adapted 
strange quark mass and the different regularization scheme are very small.

We determine the pseudocritical temperature of the chiral crossover from the inflection 
point of the subtracted quark condensate with temperature, i.e.,
\beq
	\label{eq:Tc_definition}
	\Tc
	=
	\argmax_T
	\left\lvert
	\frac{\partial \Delta_{\upu\ups}}{\partial \+ T}
	\right\rvert
\eeq
and find
\beq
	\label{eq:Tc_result}
	\Tc = \SI{156(1)}{\mega\eV}
\eeq
at vanishing chemical potential. The error given is purely numerical in nature.
In the left diagram of Fig.~\ref{fig:condensate_phase_diagram}, we show the subtracted quark
condensate as a function of temperature at vanishing chemical potential. As described in the
previous section, our result for the pseudocritical temperature agrees by construction with 
the lattice result. A nontrivial result, however, is the almost perfect match regarding the 
steepness of the chiral transition. Another highly nontrivial result is the matching of the 
unquenched gluon propagator \cite{Fischer:2012vc} with lattice results \cite{Aouane:2012bk}, 
as discussed and summarized in Ref.~\cite{Fischer:2018sdj}.

Our result for the phase diagram at nonzero chemical potential is shown in the right diagram of
Fig.~\ref{fig:condensate_phase_diagram}. The chiral crossover line (dashed black) becomes steeper
with increasing chemical potential and terminates in a second-order CEP at
\beq
	\label{eq:CEP_result}
	\bigl( T^\textup{CEP}, \, \muB^\textup{CEP} \bigr)
	=
	( \SI{119(2)}{}, \SI{495(2)}{} ) \, \si{\mega\eV}
\eeq
followed by the coexistence region (shaded gray) of a first-order transition bound by spinodals 
(solid black).\footnote{See Ref.~\cite{Gunkel:2019xnh} for a more detailed discussion of the
coexistence region between the spinodal lines.}
Furthermore, we show the line of baryon chemical potential to temperature ratio $\muB / \+ T = 3$
(dotted black), emphasizing that the CEP occures at rather large chemical potential with a ratio of
$\muB^\textup{CEP} / \, T^\textup{CEP} \approx 4.2$. Our updated value for the location of the 
critical endpoint is only slightly different than the previous DSE result of Ref.~\cite{Fischer:2014ata}. 
Again, the error in Eq.~\eqref{eq:CEP_result} is purely numerical. In order to estimate the systematic
error due to our truncation assumptions, we need to compare with different truncations as, e.g., employed
very recently in the framework of functional renormalization-group equations \cite{Fu:2019hdw}.
This will be a task for future work.

In the plot in Fig.~\ref{fig:condensate_phase_diagram}, we also show results for the chiral transition
obtained on the lattice (blue band) \cite{Bellwied:2015rza} (see also Ref.~\cite{Bazavov:2018mes}).
As can be seen in the plot, this band features a (very) small error at small chemical potential which 
rapidly increases toward larger chemical potential. At about $\muB / \+ T \approx 3$, the errors become so 
large that further extrapolation becomes meaningless. Combined evidence of different methods on 
the lattice points toward no critical endpoint for $\muB / \+ T \leq 2 - 2.5$
\cite{Bellwied:2015rza,Bazavov:2017dus} in agreement with our result. Furthermore, we also show results
for the freeze-out points extracted from heavy-ion collisions by different groups/methods
\cite{Alba:2014eba,Becattini:2016xct,Vovchenko:2015idt,Adamczyk:2017iwn,Andronic:2016nof,Andronic:2017pug}.
In the crossover region at small chemical potential, the different results for the freeze-out points 
spread over almost \SI{20}{\mega\eV} in temperature at and around the pseudocritical temperature extracted
on the  lattice (see \cite{Bazavov:2012vg,Borsanyi:2013hza} for a direct
comparison of lattice results and experimental data). While in the presence
of a (first-order) phase transition one would expect the freeze-out to occur at temperatures below the 
critical one, this notion is hard to formulate in the crossover region, where no unique definition
of a critical temperature exists. At large chemical potential, however, where we see a first-order transition
in our DSE results, we have to expect corrections either to the location of the experimental freeze-out points 
or to the DSE results in order to account for a proper ordering of temperatures. In this respect we would like 
to point out that potential corrections to the DSE calculation have already been identified (on a qualitative 
basis), which have the potential to shift the CEP to larger temperatures and/or chemical potentials thereby 
resolving this problem \cite{Eichmann:2015kfa}.   

\subsection{\label{sec:results:chi_u}Quark number fluctuations}

\begin{figure*}[t]
	\centering%
	\includegraphics[trim=0 2mm 0 0]{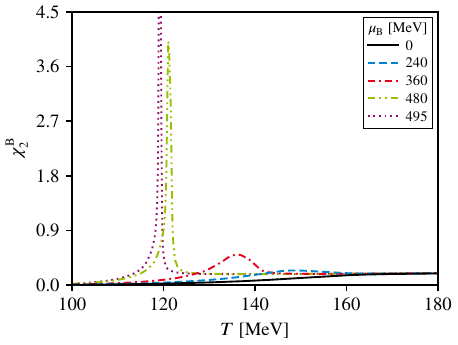}%
	\hspace*{2em}%
	\includegraphics[trim=0 2mm 0 0]{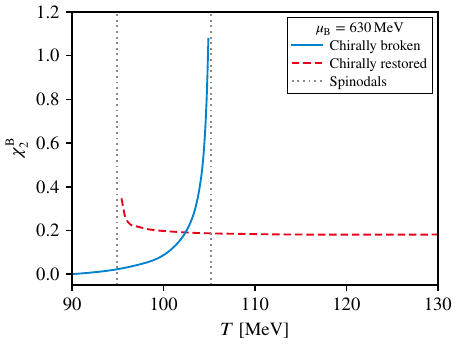}%
	\caption{\label{fig:chi_2_B}%
		Second-order baryon number fluctuation approaching the critical endpoint
		(left) and beyond in the first-order region of the phase diagram (right).
	}%
\end{figure*}

After the discussion of the phase diagram in the last subsection, we now focus on our results for the fluctuations.
In the left diagram of Fig.~\ref{fig:chi_2_u}, the second-order up/down quark fluctuation is shown as a function of 
temperature at vanishing chemical potential (solid black) and compared to results from lattice QCD
\cite{Bazavov:2011nk,Borsanyi:2011sw}. The agreement between both approaches is not as good as for 
the quark condensate but still very reasonable. The DSE result increases up to
temperatures of $T \approx \SI{165}{\mega\eV}$ and reaches an asymptotic value of approximately $0.72$ 
in the high-temperature region. Clearly, this saturation is below the Stefan--Boltzmann limit
($\chi_2^\upu \to 1$ as $T \to \infty$) and happens at much too low temperatures. We attribute
this to a known deficiency of our quark-gluon interaction.\footnote{Due to numerical efficiency, our vertex 
ansatz Eq.~\eqref{eq:vertex_ansatz} takes only the leading Dirac structure $\gamma_\nu$ into account.
In the Landau gauge employed in this work, the full vertex contains 24 different structures. Half 
of these are only present when chiral symmetry is broken, i.e., these terms react strongly on the chiral
restoration around $\Tc$. This effect is not captured by the ansatz. 
As a result, the continuous weakening of the quark-gluon interaction that drives the system and its
fluctuations toward the Stefan--Boltzmann limit is not properly represented, thus leading to the 
high-temperature artifacts seen in the left diagram of Fig.~\ref{fig:chi_2_u}. In principle, this behavior 
could be mimicked by making the vertex strength parameter temperature dependent, i.e.,
$d_1 = d_1(T)$ \cite{Mueller:2010ah}. Such a modification could be motivated and guided, e.g., by an
explicit calculation of (parts of) the vertex as a function of temperature. Preliminary results of
this endeavor have been discussed in Refs.~\cite{Welzbacher:2016,Contant:2018zpi} and need to be corroborated.}
Most important for the purpose of this work, however, is the temperature region $120$ -- $\SI{160}{\mega\eV}$
below and around the crossover temperature, where the vertex ansatz delivers satisfying results.  

Since $\chi_2^\upu$ experiences the most rapid growth in the region of the chiral crossover,
its inflection point with temperature can also be used to define the pseudocritical temperature. We
find $\Tc^{\,(\chi_2)} = \SI{153}{\mega\eV}$, which is only slightly lower than the value from the
inflection point of the subtracted condensate determined in the previous section to
\SI{156}{\mega\eV}. Note again that there is no unique definition of the critical temperature due to the
crossover nature of the transition and both values are in agreement with lattice QCD
\cite{Borsanyi:2010bp,Bazavov:2011nk,Endrodi:2011gv,Bellwied:2015rza,Bazavov:2018mes}.
While we do not expect $\chi_2^\upu$ to be strongly affected by our choice of $\mus=0$ (instead of implementing strangeness neutrality), this is certainly different for $\chi_2^\ups$. We therefore
postpone a comparison of this quantity with corresponding lattice results to future work.

Next, we turn to nonzero chemical potential. The right diagram of Fig.~\ref{fig:chi_2_u} displays the up/down quark number
density $n_\upu = T^3 \+ \chi_1^\upu$ as a function of temperature for three different
chemical potentials around our CEP, Eq.~\eqref{eq:CEP_result}. For $\muu = \muB^\textup{CEP} / \, 3 = \SI{165}{\mega\eV}$
(dashed red), the slope tends to infinity at $T = T^\textup{CEP}$ corresponding to a 
diverging second-order fluctuation. For chemical potentials above the critical one, the system
undergoes a first-order phase transition. Thus, the density is discontinuous across the phase boundary
and shows a finite jump (solid blue). This behavior is consistent with results obtained in
effective models (see, e.g., Refs.~\cite{Buballa:2003qv,Schaefer:2006ds}).
The location of the phase transition lies within the (at this chemical potential small) region between the upper and 
lower spinodal line shown in our phase diagram (cf.~Fig.~\ref{fig:condensate_phase_diagram}).
Below the critical chemical potential, where the transition
is an analytic crossover (dash-dotted green), the slope is finite around the pseudocritical temperature, and
the density changes continuously as a function of temperature for all $\muu < \muu^\textup{CEP}$.
In general, at large temperatures the density is an almost linear function of the temperature regardless
of the value of the chemical potential.

\subsection{\label{sec:results:chi_B}Baryon number fluctuations}

\begin{figure*}[t]
	\centering%
	\includegraphics[trim=0 2mm 0 0]{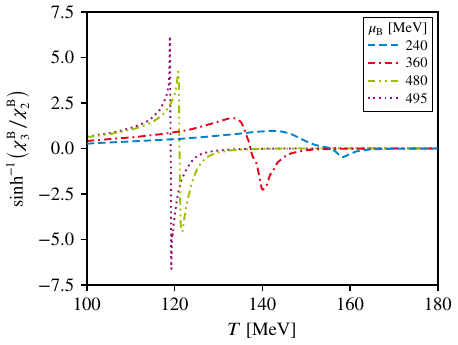}%
	\hspace*{2em}%
	\includegraphics[trim=0 2mm 0 0]{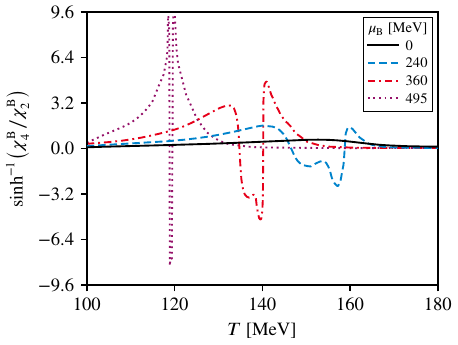}%
	\caption{\label{fig:chi_32_42_B}%
		Skewness ratio $\chi^\upB_3 / \chi^\upB_2$ (left) and kurtosis ratio $\chi^\upB_4 / \chi^\upB_2$
		(right) approaching the critical endpoint. We show the inverse hyperbolic sine
		of the ratios for better visibility. Since $\sinh^{-1}(x) \sim x$ as $x \to 0$ and
		$\sinh^{-1}(x) \sim \pm \log(2 \+ \lvert x \rvert)$ as $x \to \pm \infty$, one gets in principle
		a logarithmic plot in both positive and negative direction.
	}%
\end{figure*}

Having the quark number fluctuations at hand, we are now able to compute the baryon number
fluctuations. In particular we are interested in the changes induced by growing chemical potential 
in various ratios of baryon number fluctuations as we approach the critical endpoint. 
In the present work, we restrict ourselves to quark number fluctuations diagonal in quark flavor and neglect
off-diagonal elements that are much harder to be determined and are relegated to future 
work.\footnote{This is justified by lattice results indicating that off-diagonal 
correlations are subleading as compared to diagonal ones \cite{Bellwied:2015lba}.}
Then, the $n$\textsuperscript{th}-order baryon number fluctuation is given by
\beq
	\chi_n^\upB = \frac{1}{3^n} \, \bigl( 2 \+ \chi_n^\upu + \chi_n^\ups \bigr)
\eeq
with $n \geq 1$. 
We choose a selection of fixed baryon chemical potentials and evaluate the fluctuations as a function
of temperature. The results are shown in the left diagram of Fig.~\ref{fig:chi_2_B}, where we display the
second-order baryon number fluctuation approaching the CEP. At $\muB = \SI{0}{\mega\eV}$ (solid black),
the behavior is similar to the up quark number fluctuation, cf.~Fig.~\ref{fig:chi_2_u}, i.e., we find
a monotonous increase below and around the crossover transition. At nonzero chemical potential about 
halfway toward the critical endpoint (dashed blue), a bulge begins to develop at and around the 
pseudocritical transition temperature. This bulge becomes larger as we further increase the chemical
potential  (dash-dotted red). Close to the location of the CEP, the bulge grows considerably and becomes
a sharp peak (dash-dot-dotted green) which finally diverges at the CEP (dotted purple), 
as expected from the behavior of the quark number density, discussed above.\footnote{While in principle 
one could fine-tune the chemical potential to come arbitrarily close to the actually divergence, 
in practice limited numerical accuracy together with finite computer resources always lead to a very 
large but still finite correlation length.} 

The behavior of $\chi_2^\upB$ in the first-order region of the phase diagram is shown in the right
diagram of Fig.~\ref{fig:chi_2_B}. The second-order baryon number fluctuation shows 
two branches corresponding to the chirally broken solution (solid blue) and partially chirally restored solution
(dashed red) of the DSE for the quark propagator. The overlap of the two solutions defines the coexistence
region of the first-order transition that is bounded by the spinodals at temperatures indicated by vertical 
dotted, gray lines. For temperatures above and below the coexistence region, $\chi_2^\upB$ is only very slowly 
varying with temperature.

Next, we discuss ratios of fluctuations that are directly related to experimental quantities in
heavy-ion collisions through event-by-event analyses (see Eq.~\eqref{eq:chi_ratios}). 
In Fig.~\ref{fig:chi_32_42_B}, we plot the skewness ratio $\chi^\upB_3 / \chi^\upB_2$ (left) and the
kurtosis ratio $\chi^\upB_4/\chi^\upB_2$ (right) again as a function
of temperature for various lines of constant chemical potential up to the critical endpoint. These show
distinctive features. Whereas for small chemical potential up to halfway toward the critical endpoint
all structures are very small in size, these grow rapidly when the CEP is approached. The skewness develops
a characteristic rise with temperature accompanied by a zero crossing and subsequent equally drastic 
decrease in magnitude when the temperature is further increased. This structure becomes extremely pronounced
close to the CEP. Correspondingly, the kurtosis ratio $\chi^\upB_4 / \chi^\upB_2$ develops an asymmetric double-peak structure across the phase boundary.

There are a number of caveats when comparing results from theoretical calculations with data extracted 
from experiment. These are related to the experimental conditions such as the finite volume and the finite
temporal extent of the fireball and the question whether and when the system is in thermodynamical equilibrium.
Furthermore, these are related to details of the experimental analysis such as centrality cuts, the question
whether proton number fluctuations are a proxy for baryon number fluctuations, and potential other issues, 
see the reviews \cite{Luo:2017faz,Bzdak:2019pkr} and references therein. Still, there is considerable 
interest in comparing experimental data with results from theoretical calculations along the phase boundary.  
Such a comparison is done in the following. 

\begin{figure}[t]
	\centering%
	\includegraphics[trim=0 2mm 0 0]{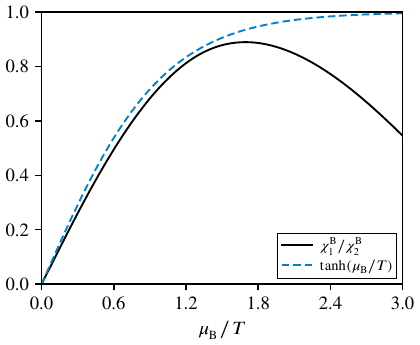}%
	\caption{\label{fig:chi_12_B_tanh}%
		The ratio $\chi^\upB_1/\chi^\upB_2$ as a function of $\muB / \+ T$ compared to the HRG
		result $\tanh(\muB / \+ T)$ \cite{Karsch:2003zq,Karsch:2010ck} along the crossover line.
	}
\end{figure}

In Fig.~\ref{fig:chi_12_B_tanh}, we display our results for the ratio $\chi^\upB_1 / \chi^\upB_2$
extracted along our crossover transition line.
For small chemical potential, i.e., up to $\muB / \+ T \lesssim 1.5$, it is expected from the 
HRG model \cite{Karsch:2003zq,Karsch:2010ck} that the ratio is approximately given 
by $\tanh(\muB / \+ T)$. This has been seen as well in the PQM model \cite{Almasi:2017bhq} and also shows
up in our calculation. Sizeable deviations only occur for larger chemical potential: After the maximum 
at $\muB / \+ T \approx 1.6$, the ratio goes down again and signals the approach to the CEP due the increase of $\chi^\upB_2$ already seen in Fig.~\ref{fig:chi_2_B}. 

\begin{figure}[t]
	\centering%
	\includegraphics{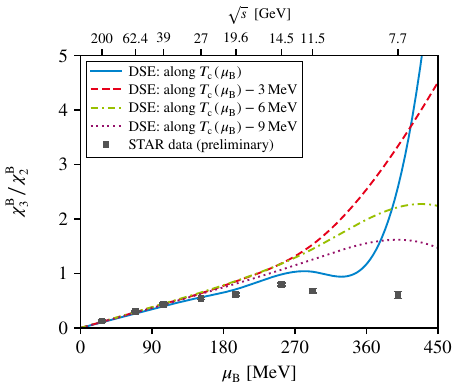}%
	\\[1mm]%
	\includegraphics[trim=0 2mm 0 0]{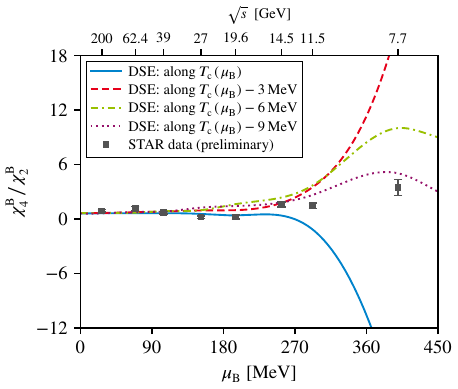}
	\caption{\label{fig:chi_32_42_B_comparison}%
		Skewness ratio $\chi^\upB_3 / \chi^\upB_2$ (top) and kurtosis ratio
		$\chi^\upB_4 / \chi^\upB_2$ (bottom) along the crossover line and for lines with a fixed
		temperature distance from the crossover. Also shown are preliminary data from the STAR
		collaboration \cite{Luo:2015ewa,Luo:2015doi} at most central collisions.
		We adopt the $\muB$-$\sqrt{s}$ translation from Ref.~\citep{Adamczyk:2017iwn}.
	}
\end{figure}

Even more interesting are the ratios involving higher-order fluctuations. 
In Fig.~\ref{fig:chi_32_42_B_comparison}, we present results for the skewness ratio
$\chi^\upB_3 / \chi^\upB_2$ (upper diagram; blue, solid line) and the kurtosis ratio
$\chi^\upB_4 / \chi^\upB_2$ (lower diagram; blue, solid line) along our chiral phase boundary determined
from the inflection point of the chiral condensate, Eq.~\eqref{eq:Tc_definition}. For the skewness,
this criterion leaves us on the left and positive branch of the oscillations shown in
Fig.~\ref{fig:chi_2_B}. For the kurtosis, however, we probe the (small) negative region around the
phase boundary once the chemical potential becomes large. At small chemical potential, there is very
good agreement between our results and the (preliminary) data from the STAR collaboration.
From $\sqrt{s} = \SI{14.5}{\giga\eV}$ on, about halfway toward our CEP, this agreement becomes worse
and disappears for $\sqrt{s} \leq \SI{11.5}{\giga\eV}$.
In order to discuss this aspect further, we also evaluated the skewness and kurtosis ratios on lines
with a fixed temperature  distance of $3$, $6$, and $\SI{9}{\mega\eV}$ below the crossover line.
The general idea of this comparison is to study the impact of two different effects: (i) as mentioned
already several times, there is no unique definition of the critical temperature in the crossover region
and it is therefore by now means clear, whether a given definition should coincide with the experimental
freeze-out line or not; (ii) as the chemical potential becomes larger and a potential CEP is approached,
it is also not clear whether the freeze-out line and the crossover line have the same curvature.
In other words, it may very well be, that the freeze-out line bends stronger than the crossover line and the
distance between the two lines grows with chemical potential. 

Taken at face value, our results shown in Fig.~\ref{fig:chi_32_42_B_comparison} seem to support
this notion at least on a qualitative level. At small chemical potential, the variations in both
ratios with temperature are very small and cannot be discriminated by the data. The two data points
at $\sqrt{s}=\SI{19.6}{\GeV}$ and $\sqrt{s}=\SI{14.5}{\GeV}$, however, favor a scenario with a freeze-out line very 
close to the crossover line, and we conclude that this is generally the case for $\sqrt{s}>\SI{14.5}{\GeV}$. 
The results for the kurtosis ratio at $\sqrt{s}=\SI{11.5}{\GeV}$ and $\sqrt{s}=\SI{7.7}{\GeV}$, however,
suggest that the freeze-out line in this region of the phase diagram is separated from the
crossover line by at least $\SI{9}{\MeV}$.
The corresponding results for the skewness ratio show the same general trend, although
on a less quantitative level than the ones for the kurtosis. 

There are several caveats involved in the comparison of the experimental STAR data and our results 
in Fig.~\ref{fig:chi_32_42_B_comparison}. Some caveats on the experimental side have been discussed already above 
and are reviewed in Refs.~\cite{Luo:2017faz,Bzdak:2019pkr}. Our theoretical calculation suffers from several
limitations. First, we did not yet take into account the effect of off-diagonal contributions to the
baryon number fluctuations. Second, there may be a substantial error associated with the precise location
of the critical endpoint. The source of this error is entirely located in the
truncation for the quark-gluon vertex and may be reduced in the future by extended DSE calculations 
\cite{Eichmann:2015kfa,Contant:2018zpi} and/or systematic comparisons with similar calculations in the 
functional renormalization group framework \cite{Braun:2009gm,Braun:2014ata,Fu:2019hdw}.
Third, one has to bear in mind that the fluctuations
triggering the CEP in this work are gluonic in nature. Consequently, the critical exponents of our CEP
are mean field. In Ref.~\cite{Fischer:2011pk}, it has been shown that the inclusion of fluctuations from 
composite pion and sigma fields in the quark DSE serves to generate the critical $\OO(4)$ physics of the
chiral two-flavor theory. We therefore expect that the extension of that framework toward chemical
potential places our CEP in the correct $\ZZ(2)$ universality class due to the fluctuating sigma field.
Furthermore, one may expect a decrease of the size of the critical region around the CEP \cite{Schaefer:2006ds},
which in turn will drive the results shown in Fig.~\ref{fig:chi_32_42_B_comparison} further toward 
the STAR data even if the location of the CEP remains unchanged. Since the inclusion of pions and the 
sigma leads to a significant increase in complexity and CPU time in our calculations, this is
left for future work. A first step toward this can be found in Ref.~\cite{Gunkel:2019xnh}.

\section{\label{sec:summary}Summary and conclusions}

In this work, we extracted ratios of cumulants involving the skewness and the kurtosis 
from baryon number fluctuations at nonzero temperature and chemical potential. To this end,
we employed a framework of Dyson--Schwinger equations for $\Nf=2+1$ quark flavors, which 
has been studied extensively in the past \cite{Fischer:2018sdj} and shown to agree with 
lattice results in the small and moderate chemical-potential region. At large chemical 
potential, where lattice QCD cannot be applied, this approach features a critical
endpoint at $\bigl( T^\textup{CEP}, \, \muB^\textup{CEP} \bigr) = (119,495) \, \si{\mega\eV}$. Due to inherent 
limitations of the truncation scheme used, this value may have systematic errors of at least 
twenty percent. Future heavy-ion collision experiments such as FAIR/CBM, NICA, and the STAR Fixed-Target
program will be able to probe the corresponding region of the QCD phase diagram. In order to facilitate 
these experiments and to make contact with already existing preliminary data from the BES at RHIC,
we determined skewness and kurtosis ratios along our crossover line
up to the CEP. Furthermore, we scanned lines of equal temperature distance 
below the transition line. For chemical potentials $\muB < \SI{250}{\mega\eV}$, our results are in
agreement with the STAR data. For larger values we obtain qualitative and quantitative 
differences when we approach the CEP on the crossover transition line. However, qualitative 
agreement between our results and the STAR data can be obtained if we assume that the
freeze-out line and the transition line separate at larger chemical potential. We also
discussed several caveats in this interpretation, which need to be checked in future work.   


\begin{acknowledgments}
	We thank Bernd-Jochen Schaefer for many stimulating discussions on the physics of fluctuations.
	Furthermore, we thank Thorsten Steinert, Richard Williams, and Fei Gao for fruitful discussions.
	This work has been supported by the Helmholtz Graduate School for Hadron and Ion Research for FAIR,
	the GSI Helmholtzzentrum f\"{u}r Schwerionenforschung, the Helmholtz International Center for FAIR
	within the LOEWE program of the State of Hesse, and the BMBF under contract 05P18RGFCA.
	M.B.~acknowledges support by the Deutsche Forschungsgemeinschaft (DFG, German Research Foundation)
	through the CRC-TR 211 ``Strong-interaction matter under extreme conditions''---project number
	315477589---TRR 211. Feynman diagrams were drawn with \textit{JaxoDraw} \cite{Binosi:2008ig}.
\end{acknowledgments}


\bibliography{FluctuationsBibliography}

\end{document}